\documentclass[structabstract]{aa}  
\usepackage{graphicx}
\usepackage{epstopdf}
\usepackage{ulem}
\usepackage{txfonts}
\usepackage{natbib}
\begin{document}

   \title{The extent of mixing in stellar interiors: the open clusters Collinder\,261 and 
   Melotte\,66\thanks{Based on observations collected at ESO telescopes under 
Guaranteed Time Observation programmes 071.D-0065, 072.D-0019, and  076.D-0220.}
}

 \author{
          Arnas Drazdauskas\inst{1},
          Gra\v{z}ina Tautvai\v{s}ien\.{e}\inst{1},
          Sofia Randich\inst{2},   
          Angela Bragaglia\inst{3},\\
           \v{S}ar\=unas Mikolaitis\inst{1},   
     \and
  Rimvydas Janulis\inst{1}         
 }
   \institute{Institute of Theoretical Physics and Astronomy, Vilnius University, Saul\.{e}tekio al. 3, LT-10222, Vilnius,
Lithuania \\
              \email{arnas.drazdauskas@tfai.vu.lt }
         \and
         INAF - Osservatorio Astrofisico di Arcetri - Largo Enrico Fermi 5, I-50125 Firenze, Italy\\
              \email{randich@arcetri.astro.it}
         \and
            INAF - Osservatorio Astronomico di Bologna, Via Ranzani 1, I-40127 Bologna, Italy\\
              \email{angela.bragaglia@oabo.inaf.it}
             }

\authorrunning{Arnas Drazdauskas et al.}
\titlerunning{The extent of mixing in stellar interiors: the open clusters Collinder\,261 and 
   Melotte\,66}

 
  \abstract
   {Determining  carbon and nitrogen abundances in red giants provides useful diagnostics to test mixing processes in stellar atmospheres.}
   {Our main aim is to determine  carbon-to-nitrogen and  carbon isotope ratios for evolved giants in the open clusters 
Collinder\,261 and Melotte\,66 and  to compare  the results with predictions of theoretical models.}
   {High-resolution spectra were analysed using a differential model atmosphere method. 
Abundances of carbon were derived using the ${\rm C}_2$ Swan (0,1) band head at
5635.5~{\AA}. The wavelength interval 7940--8130~{\AA}, which contains CN features, was analysed 
 to determine nitrogen abundances and carbon isotope ratios.  
The oxygen abundances were determined from the [O\,{\sc i}] line at 6300~{\AA}.}
   {The mean values of the elemental abundances in Collinder\,261, as determined from seven stars, are: 
${\rm [C/Fe]}=-0.23\pm0.02$ (s.d.), ${\rm [N/Fe]}=0.18\pm0.09$,  ${\rm [O/Fe]}=-0.03\pm0.07$. The mean $^{12}{\rm C}/^{13}{\rm C}$ ratio is 
$11\pm 2$, considering four red clump stars and 18 for one star above the clump. The mean C/N ratios are $1.60\pm 0.30$ and 1.74, respectively. 
For the five stars in Melotte\,66 we obtained: ${\rm [C/Fe]}=-0.21\pm0.07$ (s.d.), ${\rm [N/Fe]}=0.17\pm0.07$,  ${\rm [O/Fe]}=0.16\pm0.04$. 
The $^{12}{\rm C}/^{13}{\rm C}$ and C/N ratios are $8\pm 2$ and $1.67\pm 0.21$, respectively.}
  {The $^{12}{\rm C}/^{13}{\rm C}$ and C/N ratios of stars in the investigated open clusters were compared with the ratios 
predicted by stellar evolution models. The mean values of $^{12}{\rm C}/^{13}{\rm C}$ ratios in Collinder\,261 and Melotte\,66 agree well with models of 
thermohaline-induced extra-mixing for the corresponding stellar turn-off masses of about 1.1--1.2~$M_{\odot}$.  
The mean C/N ratios are not decreased as much as predicted by the model in which the thermohaline- and rotation-induced 
extra-mixing act together. }


   \maketitle

\section{Introduction} 

Determining carbon and nitrogen abundances, together with $^{12}{\rm C}/^{13}{\rm C}$ isotopic ratios, in red giants 
provides useful diagnostic data to test mixing processes in stellar atmospheres. 
Observations provided evidence of the first dredge-up (\citealt{Iben65}), which brought the CN-processed 
 material up to the surface of low-mass stars 
when they reached the bottom of the red giant branch (RGB). Interestingly, they also indicated the presence of extra-mixing, which happens later on the 
giant branch (\citealt{Gilroy89}; \citealt{Luck94}; \citealt{Gratton00}; \citealt{Tautvaisiene00,Tautvaisiene05,Tautvaisiene10}; 
\citealt{Smiljanic09,Smiljanic16}; \citealt{Mikolaitis10,Mikolaitis11A,Mikolaitis11B,Mikolaitis12}, etc.). 

Ideas about the mechanisms of extra-mixing have been proposed by a number of theoreticians, since the standard first
dredge-up model alone was not able to account for the surface abundances
that had  been determined spectroscopically (see the reviews by \citealt{Chaname05, Charbonnel06} and the papers by \citealt{Cantiello10, 
Charbonnel10, Denissenkov10, Lagarde11, Wachlin11, Angelou12, Lagarde12, Lattanzio15}, and references therein). Theoretical models
of thermohaline- and rotation-induced extra-mixing are currently being intensively developed. The influence of thermohaline 
mixing and rotation varies in stars of different masses and metallicities, and still needs more observational data to be robustly modelled. 

The determination of CNO abundances and $^{12}{\rm C}/^{13}{\rm C}$ ratios in open clusters offers many advantages. 
We analyse stars that have the same age, metallicity,
and origin. Their distance and age, hence the mass and evolutionary status, can be determined more precisely than for 
field stars, at least in the pre-Gaia era. In clusters,  we have homogeneous star samples, with 
the same initial composition, in particular the CNO elements. Thus changes in their abundances are related to internal 
processes of stellar evolution.

In the present work, we have determined abundances of carbon,
nitrogen and oxygen, and carbon isotopic $^{12}{\rm C}/^{13}{\rm C}$ ratios  in evolved stars of the open clusters 
Collinder\,261 and Melotte\,66. The new results of these 
 two old open clusters with low turn-off masses (i.e. 1.1--1.2~$M_{\odot}$, given their old age), taken together with the results of previous studies, 
 are used to evaluate the theoretical models of extra-mixing. 

The first survey of the open cluster Collinder\,261 (Cr\,261) was made by \citet{BH1975}. They determined it was a rich 
cluster with a diameter of 8 arc minutes. 
\citet{PJ1994} provided the first CCD photometry and claimed Cr\,261 was one of the oldest clusters, although no precise age was given.
A very old age was confirmed by later studies (\citealt{Mazur95} suggested 5 Gyr, \citealt{Gozzoli96} gave a range of 7--11 Gyr, and \citealt{Bragaglia06} preferred an age
of 6 Gyr).

The low-resolution spectroscopic study by \citet{Friel95} determined a metallicity ${\rm [Fe/H]}=-0.14$, which was updated to ${\rm [Fe/H]}=-0.16 \pm 0.13$
by \citet{Friel02}, based on 21 stars. \citet {Friel03} analyse four stars with high-resolution spectroscopy and determine ${\rm [Fe/H]}=-0.22 \pm 0.05$.
The following high-resolution spectroscopic studies determine a slightly higher 
metal content. \citet {Carretta05} investigate 
six red clump and RGB stars and find  ${\rm [Fe/H]}=-0.03 \pm 0.03$ for the five warmer stars. \citet {Silva07} also determine 
 ${\rm [Fe/H]}=-0.03 \pm 0.05$. \citet {Sestito08} obtain spectra with two UVES setups (see later), but base their results
 mostly on the bluer ones. They determine ${\rm [Fe/H]}=+0.13 \pm 0.05$ for seven RGB 
and red clump stars. We use the spectra of both their setups.

The only study of CNO elements and carbon isotope ratios in Collinder\,261 was performed by \citet {Mikolaitis12}. 
Using high resolution FEROS spectra of the six stars previously investigated by \citet{Carretta05}, they determine 
a mean carbon-to-nitrogen 
ratio of $1.67 \pm 0.06$. The average value of $^{12}{\rm C}/^{13}{\rm C} =18\pm 2$ was determined in four 
giants, and $12 \pm 1$ in two clump stars. Their results agree with the theoretical cool-bottom 
processing model proposed by \citet {Boothroyd99} and with the thermohaline-induced mixing model computed 
by \citet {Charbonnel10}

The first mention of Melotte\,66 (Mel\,66) can be found in the work by \citet {Eggen62}. They performed a photographic 
 $BV$ photometry for 20 stars. \citet {King64} made a suggestion  
that this cluster is one of the oldest in our Galaxy. Various authors quote rather similar ages, ranging from 4 to 7~Gyr 
 (\citealt {Hawarden76, Twarog79, Gratton82, Friel93, Kassis97}). In the present paper we assume the age of 
 \citet{Kassis97}, i.e. 4 Gyr.

The photometric analyses of Melotte\,66 consistently provide low values for its metallicity, about $-0.5$~dex 
 (\citealt{Hawarden76}; \citealt{Dawson78}; \citealt{Geisler84}; \citet {Twarog95}). This was confirmed by 
moderate-resolution (\citealt {Friel93} determined ${\rm [Fe/H]}=-0.51 \pm 0.11$ from four giants) and by 
high-resolution spectroscopy (\citealt {Gratton94} determined ${\rm [Fe/H]}=-0.38 \pm 0.15$ from two giants).  
The most recent investigation was carried out by \citet {Sestito08} using high-resolution spectra of six giants with 
mean ${\rm [Fe/H]}=-0.33 \pm 0.03$. 

There are no previous investigations of CNO abundances or $^{12}{\rm C}/^{13}{\rm C}$ ratios for this cluster. 
In particular, we use the spectra of the five stars that are considered members and are not fast rotators.  
  
\begin{figure}
 \includegraphics[width=0.49\textwidth]{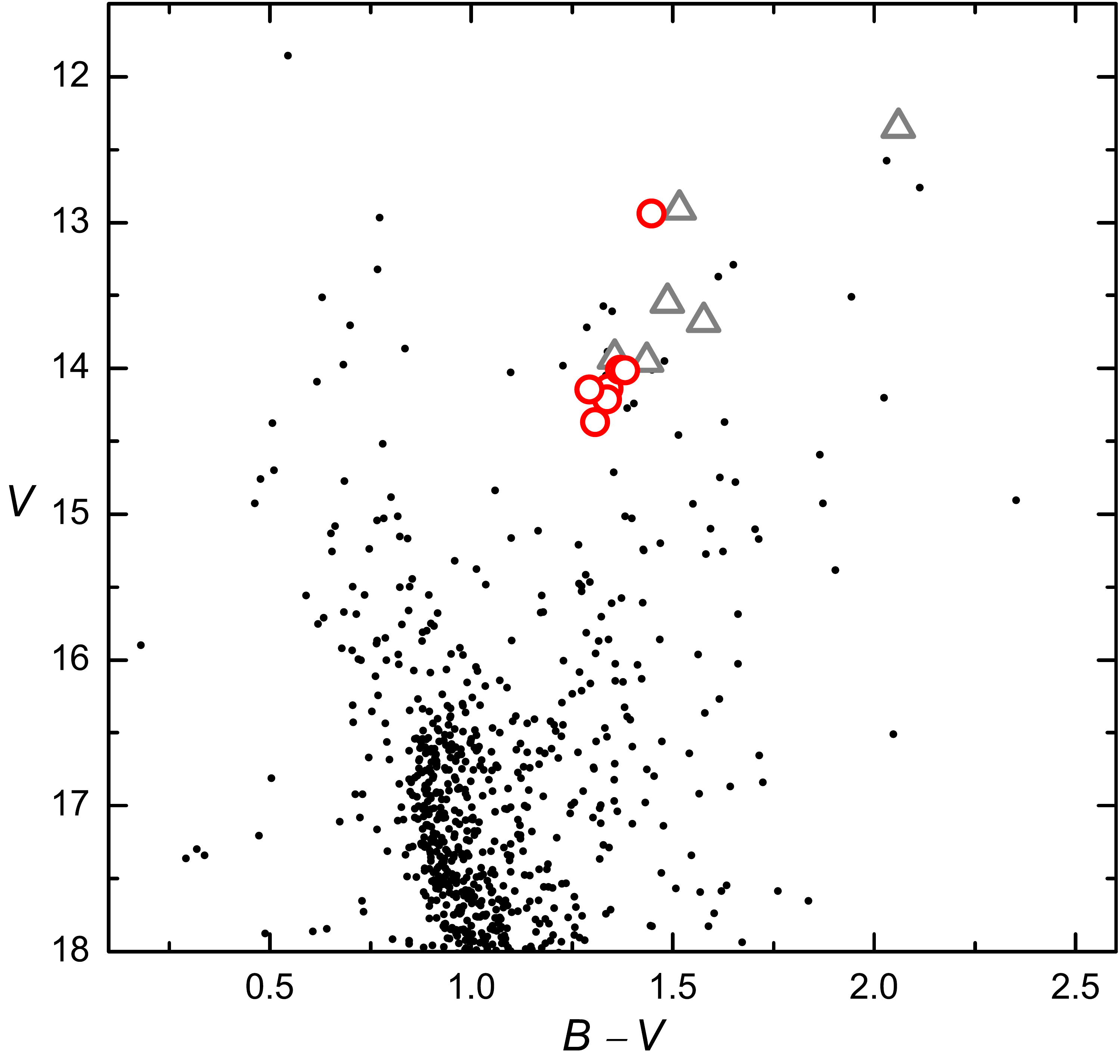}
 \caption{Colour-magnitude diagram of the open cluster Collinder\,261. The stars investigated in this work are 
indicated by red open circles, while the stars of \citet {Mikolaitis12} are indicated with open triangles. 
The diagram is based on $BVI$\ photometry by \citet {Gozzoli96}.} 
\label{Fig1}
\end{figure}

\begin{figure}
 \includegraphics[width=0.48\textwidth]{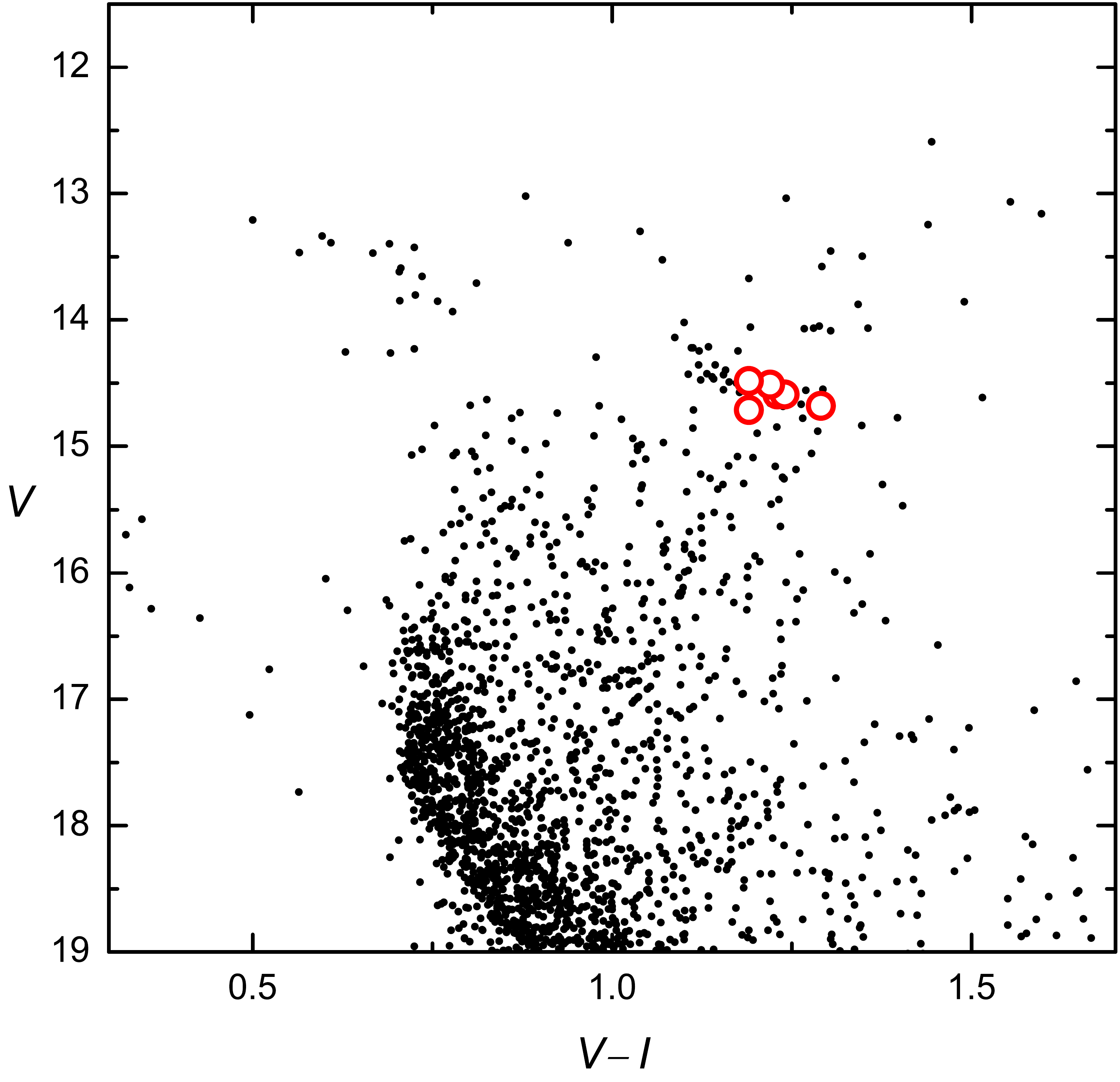}
 \caption{Colour-magnitude diagram of the open cluster Melotte\,66. The stars investigated in this work are 
indicated by red open circles. The diagram is based on $VIc$~CCD photometry by \citet {Kassis97}.} 
\label{Fig2}
\end{figure}

\section{Observations and method of analysis}

The multi-object instrument FLAMES (Fiber Large Array Multi-Element Spectrograph, \citealt{Pasquini02}) 
on the Unit 2 of the Very Large Telescope (UT2/VLT, European Southern Observatory, Chile) was used to observe 
the target clusters. 
High-resolution spectra of five giants in Mel\,66 and seven  in Cr\,261 were obtained using the UVES 
(Ultraviolet and Visual Echelle Spectrograph, \citealt{Dekker00})  with a resolving power of 47\,000. 
The clusters were observed  with both the RED580 (wavelength range $\sim4750-6800~{\AA}$) and RED860 (wavelength 
range $\sim6600-10600~{\AA}$) configurations. Details of observations and reductions are presented by \citet{Sestito08}. 

The main atmospheric parameters for the target stars were derived spectroscopically by \citet {Sestito08}.
For convenience, we present these atmospheric parameters in Table~\ref{table:1}. Two stars of Melotte\,66 were excluded 
from our analysis since one is a fast rotator and one is not a secure member, based on its radial velocity.
Figs.~\ref{Fig1} and~\ref{Fig2} show the colour-magnitude diagrams of the analysed stars in both clusters.
We analysed the spectra using a differential analysis technique. All calculations are differential with respect to 
the Sun. Solar element abundance values were taken from \citet{Asplund05}. 
Spectral synthesis was used for all abundance determinations, as well as for $^{12}{\rm C}/^{13}{\rm C}$\ ratio calculations.
The program BSYN, developed at the Uppsala University, was used for spectral syntheses. A set of plane-parallel, one
dimensional, hydrostatic, constant flux LTE model atmospheres were taken from the MARCS stellar model atmosphere and 
flux library\footnote{http://marcs.astro.uu.se/} (\citealt{Gustafsson08}). The Vienna Atomic Line Data Base (VALD, \citealt{Piskunov95}) was  
used in preparing input data for the calculations. Atomic oscillator 
strengths for the main spectral lines analysed in this study were taken from  
an inverse solar spectrum analysis (\citealt{Gurtovenko89}).

To determine the carbon abundance  in all stars we used two regions: the ${\rm C}_2$ Swan (0,1) band heads at 5135.5~{\AA} 
and 5635.2~{\AA}. 
We used the same molecular data of ${\rm C}_2$  as \citet {Gonzalez98}. 
The oxygen abundance was derived from a synthesis of the forbidden [O\,{\sc i}] line at 6300~{\AA}. The $gf$\ values for 
$^{58}{\rm Ni}$ and $^{60}{\rm Ni}$ isotopic line components, which blend with the oxygen line, were taken from
\citet{Johansson2003}. The interval 7980-8010~{\AA, which} contains strong CN features, was used to determine nitrogen 
abundance and 
$^{12}{\rm C}/^{13}{\rm C}$\ ratios. The $^{12}{\rm C}/^{13}{\rm C}$\ ratio was obtained from the $^{13}{\rm C}/^{12}{\rm N}$ 
feature at 8004.7~{\AA}. The CN molecular data for this wavelength interval were provided by Bertrand Plez.

\begin{figure}
 \includegraphics[width=0.48\textwidth]{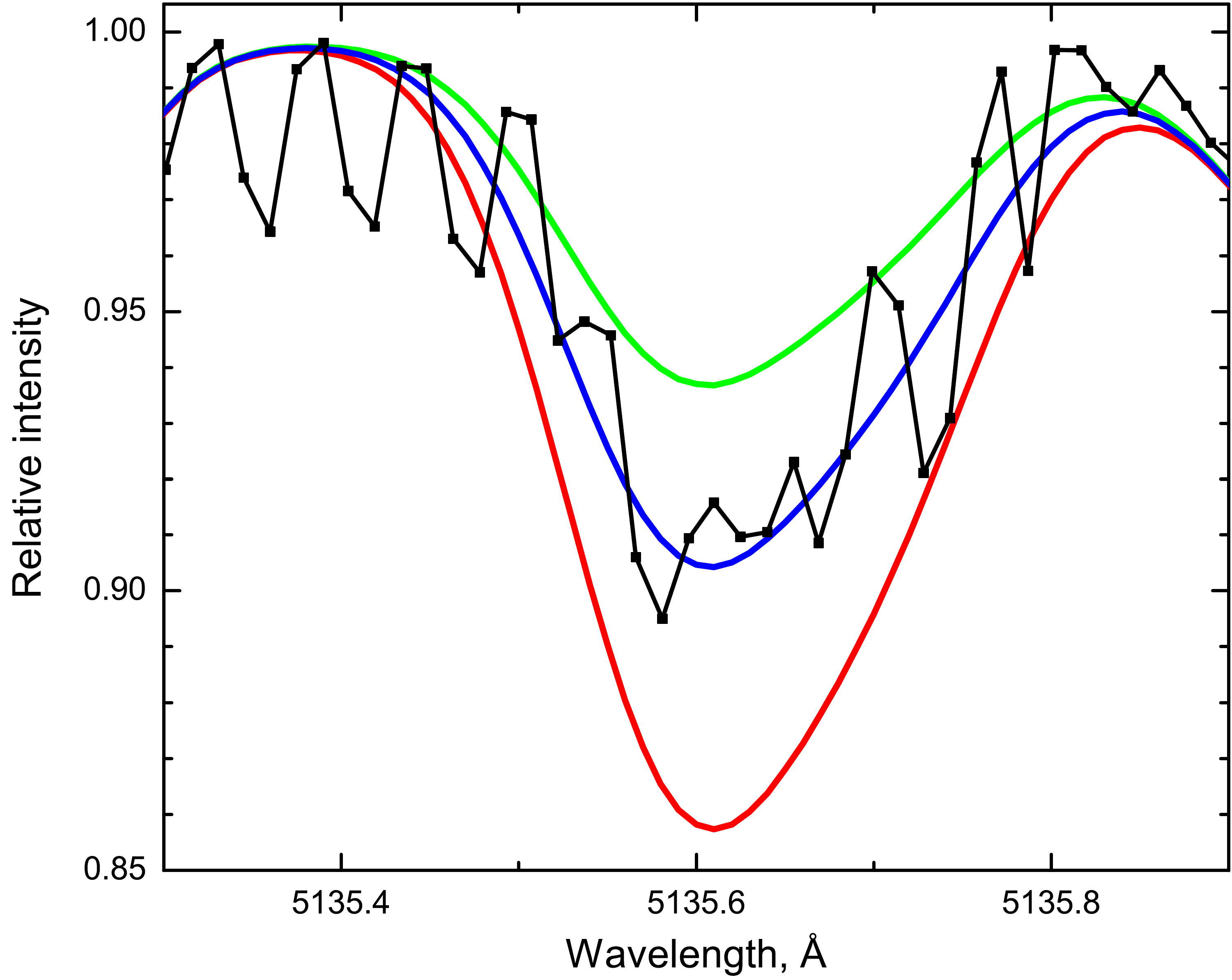}
 \caption{Fit to the ${\rm C}_2$ Swan (1,0) band head at 5135~{\AA} in the star Melotte\,66\,1785. 
The observed spectrum is shown as a black line with dots. The synthetic spectra with ${\rm [C/Fe]}=-0.52$ is shown as a blue line and  $\pm 0.1$ dex as red and green lines.} 
\label{Fig3}
\end{figure}

\begin{figure}
 \includegraphics[width=0.48\textwidth]{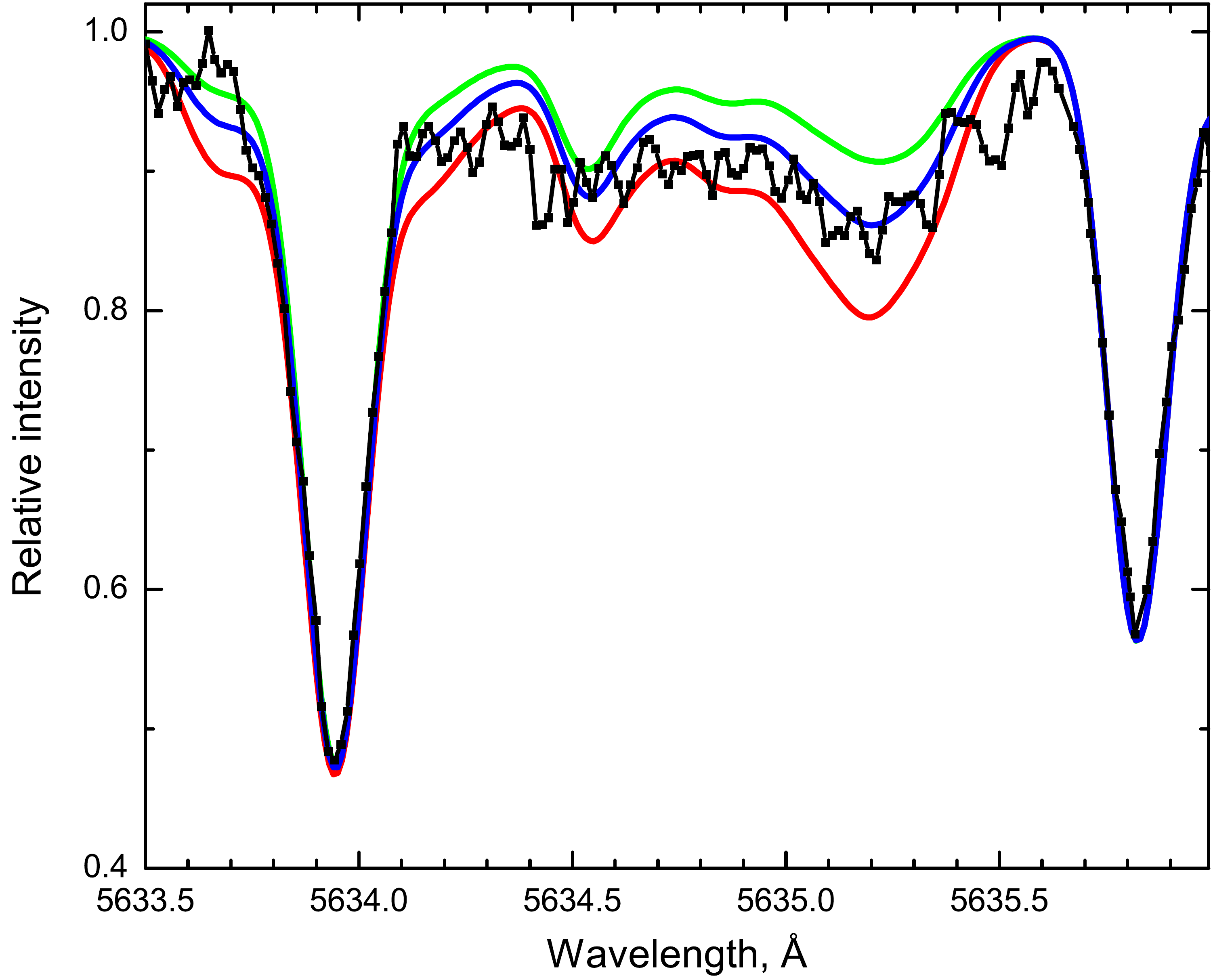}
 \caption{Fit to the ${\rm C}_2$ Swan (0,1) band head at 5635.5~{\AA} in the star Collinder\,261\,RGB05. 
The observed spectrum is shown as a black line with dots. The blue line is the synthetic spectra with ${\rm [C/Fe]}=-0.1$ and the green and red lines show the abundance of $\pm 0.1$ dex.} 
\label{Fig4}
\end{figure}

\begin{figure}
 \includegraphics[width=0.48\textwidth]{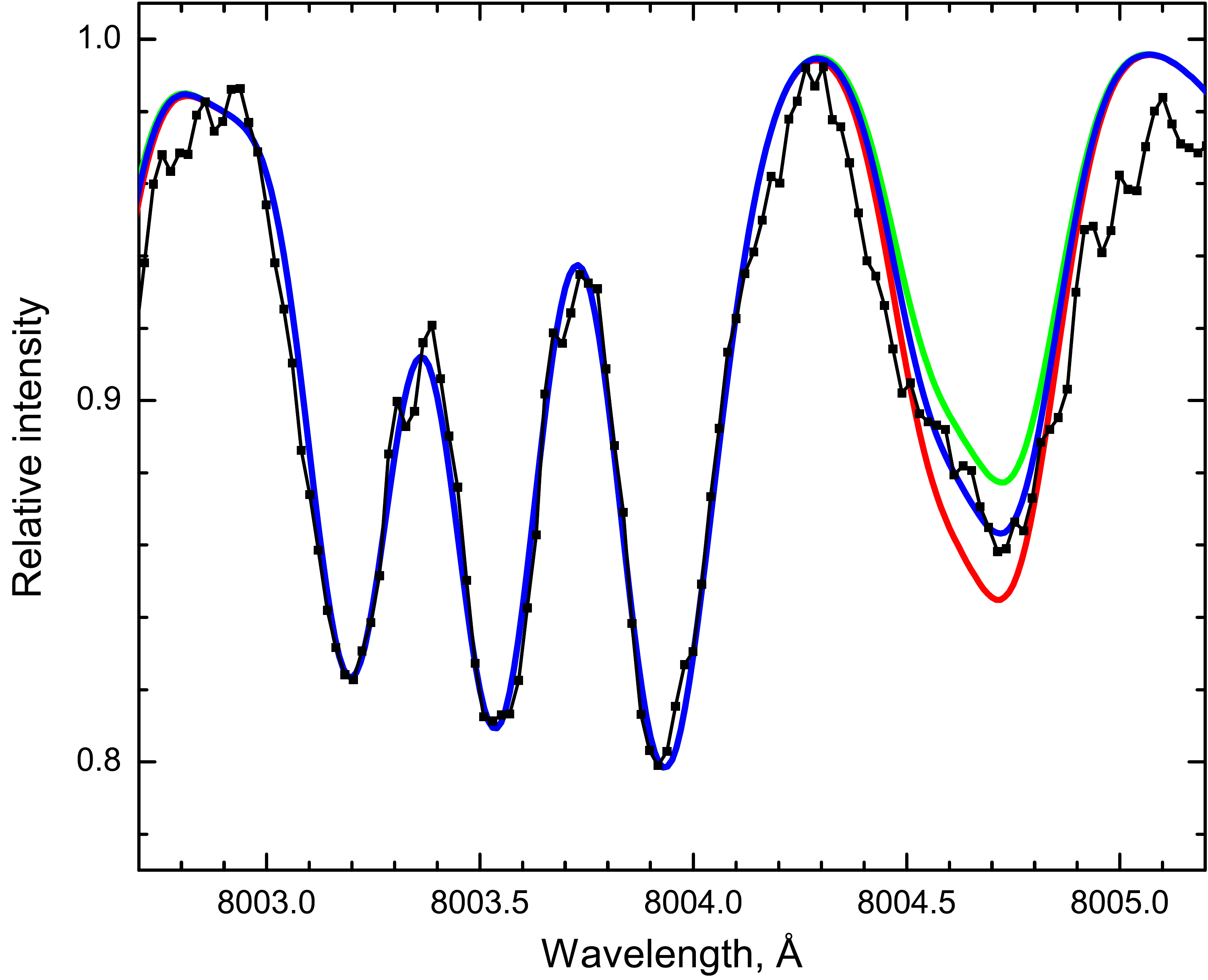}
 \caption{Fit to the CN bands in the star Melotte\,66\,1865. 
The observed spectrum is shown as a black line with dots. The synthetic spectra with ${\rm [N/Fe]}=0.42$ and 
$^{12}C/^{13}C=6$ is shown as a blue line and green and red lines represent $\pm 1$ to the $^{12}C/^{13}C$ ratio.} 
\label{Fig5}
\end{figure}

\begin{figure}
 \includegraphics[width=0.48\textwidth]{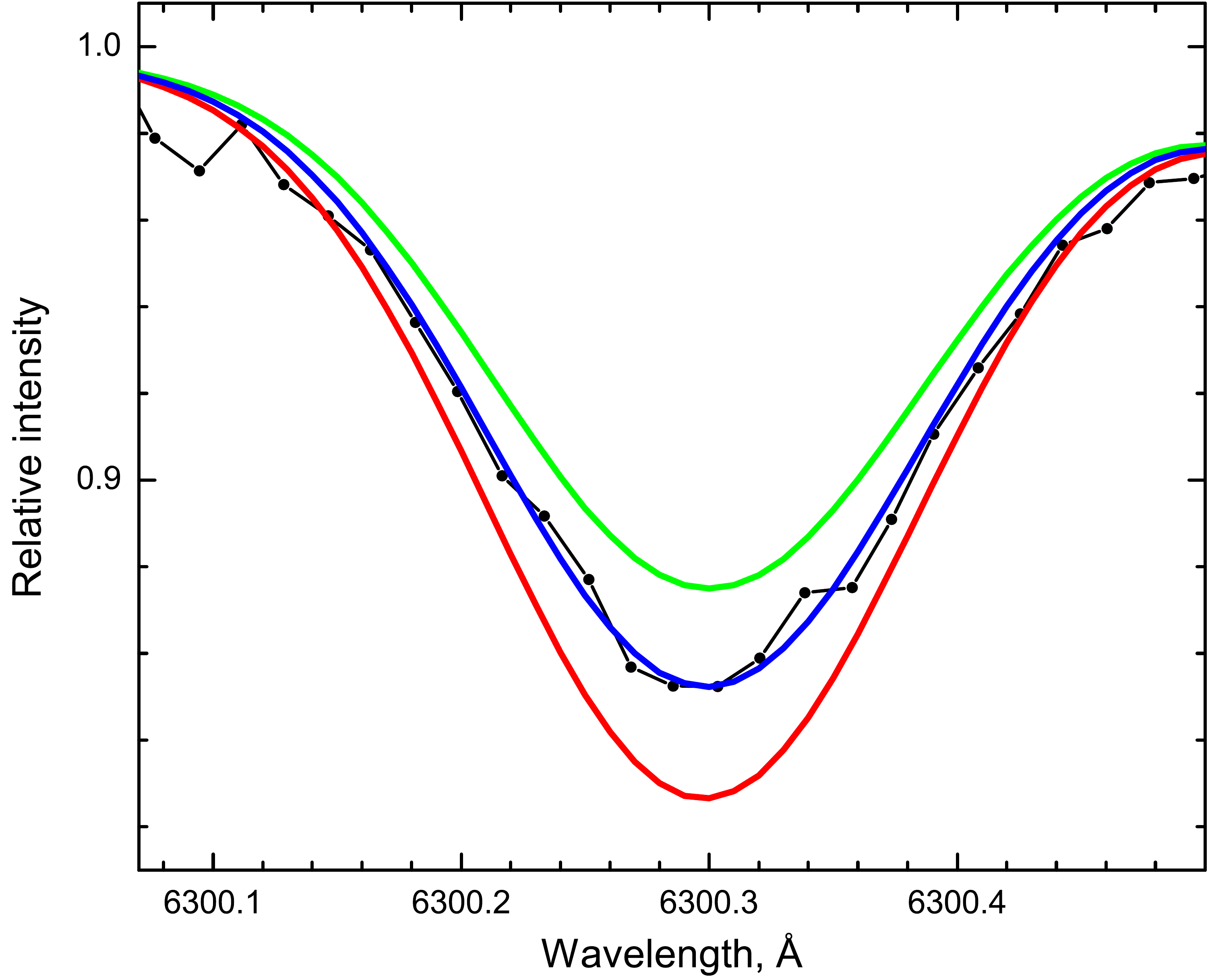}
 \caption{Fit to the forbidden [O\,{\sc i}] line at 6300.3 {\AA} in the spectrum of the star Melotte\,66\,1493. 
The observed spectrum is shown as a black line with dots. The synthetic spectra with ${\rm [O/Fe]}=0.2$ is shown as a blue line and $\pm 0.1$ dex is shown as green and red lines.} 
\label{Fig6}
\end{figure}

All the synthetic spectra were calibrated to the solar spectrum by \citet{Kurucz05} to make the analysis differential to the Sun. 
Figures~\ref{Fig3}--\ref{Fig6}  display examples of spectrum syntheses for the programme stars. The best-fit abundances were 
determined by eye.


 \begin{table}
     \centering
\begin{minipage}{250mm}
\caption{Main parameters of programme stars (\citealt {Sestito08})}
\label{table:1}
\begin{tabular}{lccccccc}
\hline\hline
\noalign{\smallskip}
 Cr~261 & $T_{\rm eff}$ & log~$g$ &$v_{\rm t}$ & [Fe/H] & $V$ & $B-V$ \\
Star  & K & & km s$^{-1}$ & & mag & mag \\
  \hline
\noalign{\smallskip}
   RGB02$^\ast$ &    4350 &   1.70 &    1.25 &    +0.12 &    12.94 &    1.45   \\
   RGB05 &    4600 &   2.00 &    1.24 &    +0.14 &    14.14 &    1.34    \\
   RGB06 &    4500 &   2.30 &    1.18 &    +0.16 &    14.01 &    1.37   \\
   RGB07 &    4546 &   2.15 &    1.20 &    +0.18 &    14.01 &    1.38   \\
   RGB09 &    4720 &   2.05 &    1.27 &    +0.04 &    14.21 &    1.34   \\
   RGB10 &    4700 &   2.35 &    1.20 &    +0.20 &    14.37 &    1.31   \\
   RGB11 &    4670 &   2.15 &    1.13 &    +0.09 &    14.15 &    1.29   \\
\noalign{\smallskip}
\hline
\noalign{\smallskip}
Mel~66 & $T_{\rm eff}$ & log~$g$ &$v_{\rm t}$ & [Fe/H] & $V$ & $V-I$  \\
Star  & K & & km s$^{-1}$ & & mag & mag \\
  \hline
\noalign{\smallskip}
\noalign{\smallskip}
1346 &    4750 &    2.00 &    1.17 &    $-0.37$ &    14.59 &         1.23 \\
1493 &    4770 &    2.15 &    1.20 &    $-0.35$ &    14.68 &         1.29 \\
1785 &    4770 &    2.05 &    1.20 &    $-0.30$ &    14.59 &         1.24 \\
1884 &    4750 &    2.45 &    1.23 &    $-0.30$ &    14.71 &         1.19 \\
2218 &    4850 &    2.39 &    1.25 &    $-0.31$ &    14.48 &         1.19 \\
\hline
\end{tabular}
\tablefoot{RGB02 is a first ascent giant. Other stars are He-core \\ burning stars. The identifications are taken from \citet {Sestito08}. }
\end{minipage}
\end{table}

Uncertainties can be divided into two categories. Firstly, there are errors that affect each line individually. 
Fitting of individual lines depends on several factors, including
uncertainties on atomic parameters, continuum placement variations, and the fitting of synthetic spectra to each line. Secondly, 
there are errors that affect all measured lines
simultaneously, such as uncertainties in the stellar atmospheric parameters used.

Table~\ref{table:2} shows a relation between the abundance estimates [El/Fe] and assumed uncertainties of the atmospheric 
 parameters in the programme star Mel\,66\,1493. 
Considering the given deviations from the  parameters used, we see that the abundances are not affected strongly. 

Since abundances of C, N, and O are also bound together by molecular equilibrium 
in the stellar atmospheres, we  investigated further how an error in one of 
them typically affects the abundance determination of another. 
Thus $\Delta{\rm [O/H]}=0.10$ causes $\Delta{\rm [C/H]}=0.04$ and $\Delta{\rm [N/H]}=0.08$;   
$\Delta{\rm [C/H]}=0.10$ causes $\Delta{\rm [N/H]}=-0.10$ and $\Delta{\rm [O/H]}=0.02$; 
$\Delta {\rm [N/H]}=0.10$ has no effect on either the carbon or the oxygen abundances.


   \begin{table}
    \centering
       \caption{Effects on derived abundances, $\Delta$[A/H], resulting from model changes
for the star Melotte\,66\,1493. }
         \label{table:2}
       \[
          \begin{tabular}{lrrccc}
             \hline
             \noalign{\smallskip}
Species & ${ \Delta T_{\rm eff} }\atop{ \pm100~{\rm~K} }$ & ${ \Delta \log g }\atop{ \pm0.3 }$ & ${ \Delta v_{\rm t} }\atop{ \pm0.3~{\rm km~s}^{-1}}$ &
             ${ \Delta {\rm [Fe/H]} }\atop{ \pm0.1}$ & Total \\
             \noalign{\smallskip}
             \hline
             \noalign{\smallskip}
C &                     0.01    &       0.08    &       0.00    &       0.02         &       0.08    \\
N &                     0.05    &       0.07    &       0.01    &       0.02         &       0.09    \\
O &                     0.01    &       0.13    &       0.00    &       0.03         &       0.13    \\
$^{12}{\rm C}/^{13}{\rm C}$ &   1       &       1       &       0       &       0         &       1.4     \\
                  \noalign{\smallskip}
             \hline
          \end{tabular}
       \]
    \end{table}


\section{Results and discussion}

The abundances of carbon, nitrogen, and oxygen relative to iron [El/Fe]\footnote{We use the customary spectroscopic notation
[X/Y]$\equiv \log_{10}(N_{\rm X}/N_{\rm Y})_{\rm star} -\log_{10}(N_{\rm X}/N_{\rm Y})_\odot$.}  
and $\sigma$ (the line-to-line 
scatter), as well as C/N and $^{12}$C/$^{13}$C ratios, are listed in Table~\ref{table:3}.

The average values and dispersions around the mean values in the six stars of Collinder\,261 are the 
 following: ${\rm [C/Fe]}=-0.23\pm0.02$ (standard deviation), ${\rm [N/Fe]}=0.18\pm0.09$,  and 
${\rm [O/Fe]}=-0.03\pm0.07$.  
From five stars in Melotte\,66, we obtained: ${\rm [C/Fe]}=-0.21\pm0.07$, 
${\rm [N/Fe]}=0.17\pm0.07$, and ${\rm [O/Fe]}=0.16\pm0.04$. 

The mean $^{12}{\rm C}/^{13}{\rm C}$ ratio in four clump stars of Collinder\,261 is $11\pm 2$ 
and 18 in the first ascent giant Cr\,261\,RGB02. The mean C/N ratio is equal to $1.60\pm 0.30$ in the clump stars and $1.74$ in Cr\,261\,RGB02. 
Our results agree well with \citet{Mikolaitis12}. Their $^{12}{\rm C}/^{13}{\rm C}$ and C/N values for two clump stars are $12 \pm 1$ and $1.67 \pm 0.06$, respectively. The mean $^{12}{\rm C}/^{13}{\rm C}$ and C/N values for the four first ascent giants in their paper is $18\pm 2$ and $1.79\pm0.16$, respectively.

Our analysis confirms that the $^{12}{\rm C}/^{13}{\rm C}$ ratio in the first ascent giants of Collinder\,261 is lower than in the more evolved clump stars that we investigated. 
The clump stars  accumulated all chemical composition changes that
took place during their evolution along the giant branch
and the helium flash.
Similar differences between first ascent giants, even though they lie above the RGB luminosity bump, 
and clump stars have been previously observed in other open clusters (M~67, \citealt{Tautvaisiene00}; NGC\,7789, \citealt{Tautvaisiene05}; NGC\,3532, \citealt{Smiljanic09}; NGC\,2506, \citealt{Mikolaitis11A}).   

The mean $^{12}{\rm C}/^{13}{\rm C}$ and C/N ratios in the clump stars of Melotte\,66 are $8\pm 2$ and $1.67\pm 0.21$, respectively. These values
can be compared to the solar value of C/N ratio which is equal to 4.07, as calculated from the solar abundances presented by 
\citet{Asplund05}, and to the $^{12}{\rm C}/^{13}{\rm C}$ ratio which is 86.8 (\citealt{Scott06}). 

The $^{12}{\rm C}/^{13}{\rm C}$ and C/N ratios of the clump stars in the two sample clusters were compared with the values predicted 
 by stellar evolution models. Among the most recent models there is the so-called thermohaline model of extra-mixing 
 (\citealt {Charbonnel10}). This model of thermohaline-instability-induced mixing is based on the ideas of \citet{Eggleton06}, 
 \citet{Ulrich72}, \citet{Charbonnel07}, and \citet{Kippenhahn80}. \citet{Eggleton06} found a mean molecular weight ($\mu$) inversion in their $1~M_{\odot}$ stellar evolution 
 model, occurring after the so-called luminosity bump on the RGB, when the hydrogen-burning shell reaches the chemically 
 homogeneous part of the envelope. The $\mu$-inversion is produced by the reaction $^3{\rm He(}^3{\rm He}, 2p)^4{\rm He}$, 
as predicted in \citet{Ulrich72}. It does not occur earlier, because the magnitude of 
the $\mu$-inversion is low and negligible compared to a stabilising $\mu$-stratification. 
\citet{Charbonnel07} computed stellar evolution models, including the ideas of \citet{Kippenhahn80}, who extended Ulrich's equations to the case of a non-perfect gas. \citet{Charbonnel07} also introduced a double 
 diffusive instability called thermohaline convection and showed its importance in the chemical evolution of red giants. This mixing connects the 
 convective envelope with the external wing of the hydrogen-burning shell and induces surface abundance modifications in evolved stars (\citealt {Charbonnel10}). 

\begin{figure}
 \includegraphics[width=0.48\textwidth]{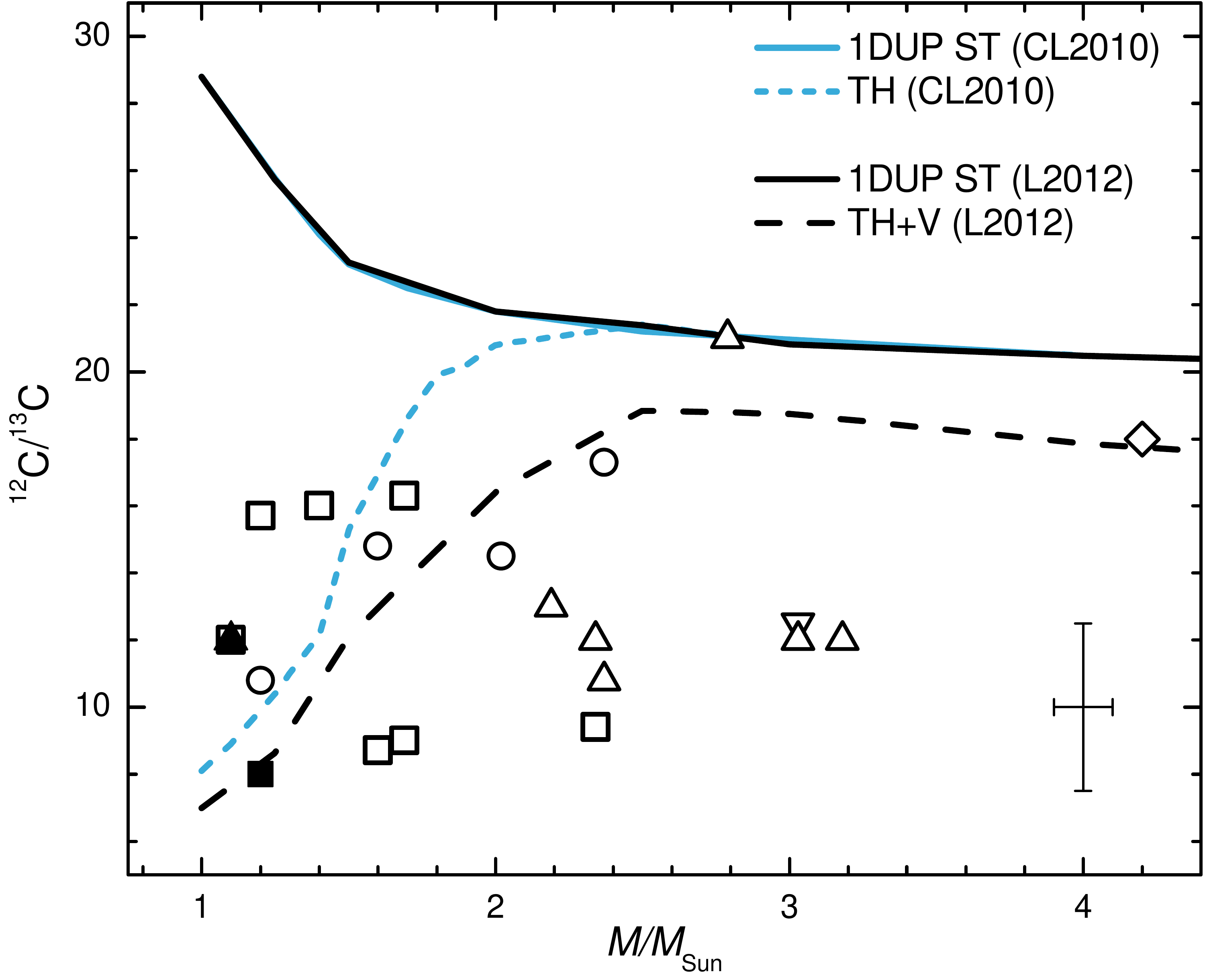}
    \caption {Average carbon isotope ratios in clump stars of open clusters
as a function of stellar  turn-off mass. 
The result for Collinder\,261 is indicated by a filled triangle and for Melotte\,66 by a filled square. 
The previous result determined for Collinder\,261 by \citet{Mikolaitis12} is indicated by an open 
square. Open squares were also used to show other results     
from \citet{Mikolaitis10, Mikolaitis11A, Mikolaitis11B, Mikolaitis12} and \citet{Tautvaisiene00, 
 Tautvaisiene05}. Results from \citet{Smiljanic09} are indicated by open triangles; from \citet{Luck94} -- 
 reversed open triangle;
from \citet{Gilroy89} -- open circles; from  \citet{KatimeSantrich13} -- open diamond.
The solid lines (1DUP ST) represent the $^{12}{\rm C}/^{13}{\rm C}$ ratios predicted for stars at the first 
dredge-up with standard stellar evolutionary models of solar metallicity by 
\citet{Charbonnel10} (blue upper line) and, more recently, \citealt{Lagarde12} (black lower 
 line). The blue dashed line (TH) shows the prediction 
when just thermohaline extra-mixing is introduced (\citealt{Charbonnel10}), and the black 
 dashed line (TH+V) is for the model that includes both 
the thermohaline- and rotation-induced mixing (\citealt{Lagarde12}),
A typical error bar is indicated (\citealt{Charbonnel10, Smiljanic09, Gilroy89}). 
}
    \label{Fig7}
  \end{figure}

\begin{figure}
 \includegraphics[width=0.48\textwidth]{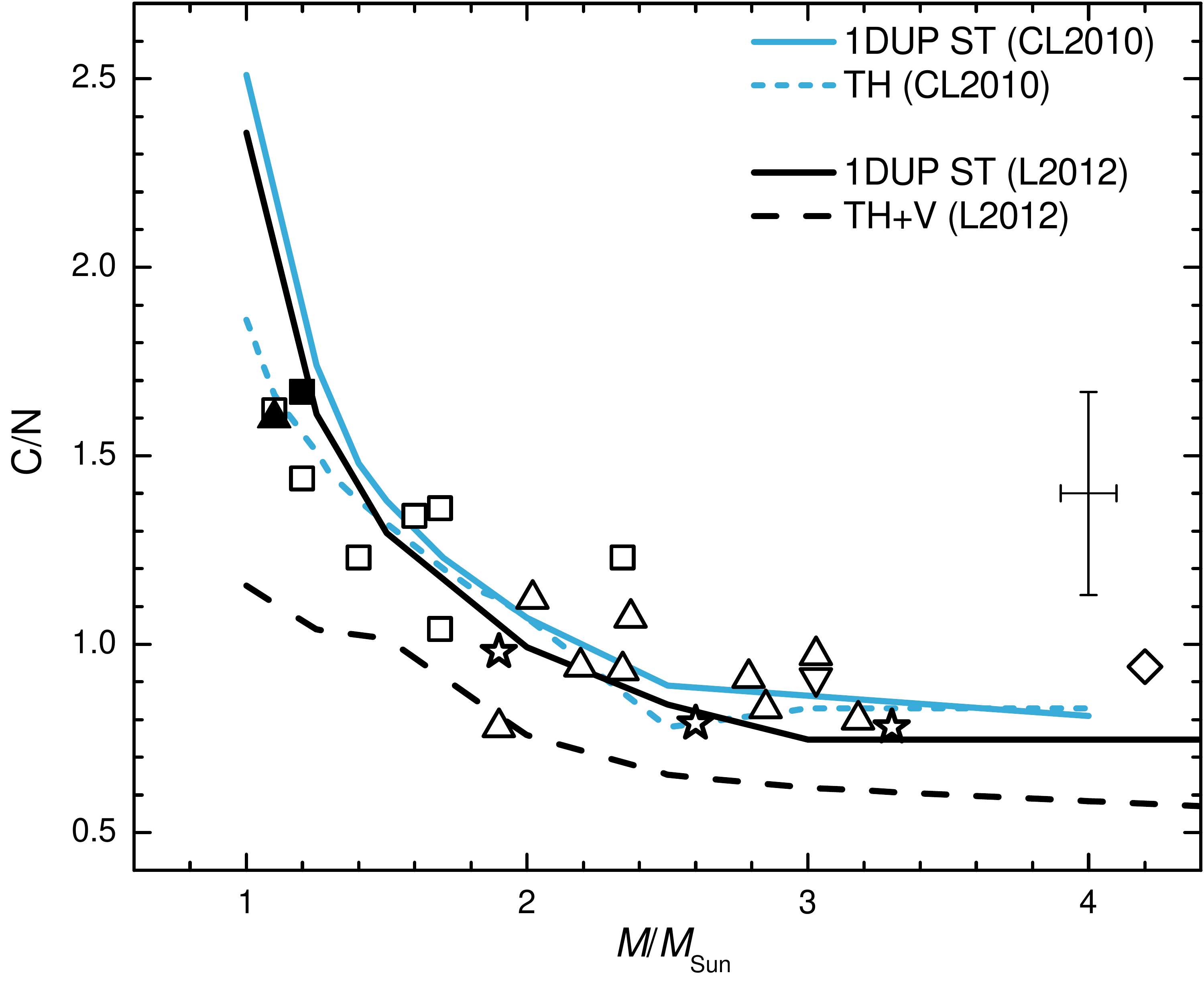}
    \caption {Average carbon-to-nitrogen ratios in clump stars of open clusters
as a function of stellar  turn-off mass. The meaning of symbols is as in Fig.~7. 
Results from \citet{Tautvaisiene15} are shown as open stars. 
 }
    \label{Fig8}
  \end{figure}

The calculation of this model was extended to metal-deficient stars by \citet{Lagarde12}, who further added the effects of rotation-induced mixing (considering the
star rotation on the zero-age main sequence). Typical initial ZAMS rotation velocities were chosen, depending on the stellar mass and based on observed rotation 
 distributions in young open clusters (\citealt{Gaige93}). The convective envelope was supposed to rotate as a solid body through the evolution. 
The transport coefficients for chemicals associated with thermohaline and rotation-induced mixing were simply added in the diffusion equation 
and the possible interactions between the two mechanisms were not considered. The rotation-induced mixing modifies the internal chemical 
 structure of main sequence stars, although its signatures are revealed only later in the stellar evolution. 

\begin{table*}
     \centering
\begin{minipage}{180mm}
\caption{Determined abundances and isotopic ratios for programme stars}
\label{table:3}
\begin{tabular}{lccccccccc}
\hline\hline
Star &  [C/Fe] & $\sigma$ [C/Fe] & [N/Fe] & $\sigma$ [N/Fe] & [O/Fe] & $\sigma$ [O/Fe] & C/N & $^{12}$C/$^{13}C$ & Evol.\\
  \hline
  \noalign{\smallskip}
  &&&& Collinder 261&&&&\\
 RGB02 & $-0.20$ & 0.04 &  0.16 &  0.02 &  $-0.10$ &  0.05 & 1.74 & 18  & g \\
 RGB05 & $-0.27$ & 0.04 & $-0.01$ &  0.05 &  $-0.09$ &  0.05 & 2.19 & 8 & c\\
 RGB06 & $-0.21$ & 0.05 &  0.29 &  0.04 &  0.09 &  0.05 & 1.26 & 13 & c \\
 RGB07 & $-0.23$ & 0.01 &  0.20 &  0.02 &  0.01 &  0.05 & 1.48 &  -- & c\\
 RGB09 & $-0.21$ & 0.07 &  0.25 &  0.02 &  $-0.04$ &  0.05 & 1.38 & 12 & c \\
 RGB10 & $-0.20$ & 0.03 &  0.17 &  0.01 &  $-0.05$ &  0.05 & 1.70 &  --  & c\\
 RGB11 & $-0.23$ & 0.01 &  0.17 &  0.03 &  $-0.11$ &  0.05 & 1.58 &  12  & c\\
\noalign{\smallskip}
\hline
\noalign{\smallskip}
Average:& $-0.23 \pm 0.02$&& $0.18 \pm 0.09$&& $-0.03 \pm 0.07$&&$1.60\pm0.30$&$11\pm2$ & \\
  \hline
\noalign{\smallskip}
 &&&& Melotte 66&&&&\\
 1346 &    $-0.32$ &    0.02 &    0.13 &    0.03 &    0.13 & 0.05    &  1.41 & 8 & c\\
 1493 &    $-0.19$ &    0.01 &    0.12 &    0.04 &    0.20 & 0.05    &  1.95 & 8 & c\\
 1785 &    $-0.23$ &    0.01 &    0.10 &    0.05 &    0.10 & 0.05    &  1.86 & 6 & c\\
 1884 &    $-0.10$ &    0.01 &    0.28 &    0.03 &    0.22 & 0.05    &  1.66 & --  & c\\
 2218 &    $-0.21$ &    0.01 &    0.23 &    0.07 &    0.14 & 0.05    &  1.45 & 10 & c\\
\noalign{\smallskip}
\hline
\noalign{\smallskip}
Average:& $-0.21 \pm 0.07$&& $0.17 \pm 0.07$&& $0.16 \pm 0.04$&&$1.67\pm0.21$&$8\pm2$ & \\
\hline
\end{tabular}
\tablefoot{Evol.: g -- first ascent giant, c -- He-core burning star. The first ascent giant RGB02 in Collinder~261 was not used for the average abundance calculations.}
\end{minipage}
\end{table*}

The mean values of $^{12}{\rm C}/^{13}{\rm C}$ and C/N ratios in clump stars are presented in Figs.~\ref{Fig7} and~\ref{Fig8}, together 
with graphical representations of the above mentioned models. 
The turn-off mass of 
 Collinder\,261 is about 1.1~$M_{\odot}$  (\citealt{Bragaglia06}) and that of Melotte\,66 about 1.2~$M_{\odot}$ (\citealt{Sestito08}). 
Results for clump stars of other open clusters  are also displayed. They were taken from  \citet{Tautvaisiene15}, \citet{KatimeSantrich13}, 
 \citet{Mikolaitis10, Mikolaitis11A, Mikolaitis11B, Mikolaitis12}, \citet{Smiljanic09}, \citet{Tautvaisiene00, Tautvaisiene05}, \citet{Luck94}, 
and \citet{Gilroy89}. To produce the average values shown in the figures, we used only red clump stars, since they provide information on the final composition
changes after all the evolution along the red giant branch.
 
The average values of $^{12}{\rm C}/^{13}{\rm C}$ ratios in clump stars of Collinder\,261 and Melotte\,66 agree well with models of 
extra-mixing (both thermohaline- and thermohaline+rotation-induced mixing since they are quite similar for the corresponding stellar 
turn-off masses of about 1.1--1.2~$M_{\odot}$). The value of Melotte\,66 is lower than that of Collinder\,261 since the metallicity of this cluster 
 is lower by about 0.5~dex.  For stars of larger turn-off masses, the theoretical extra-mixing values start approaching 
the $1^{st}$ dredge-up model, while the observational $^{12}{\rm C}/^{13}{\rm C}$ ratios remain quite low. 
 
The C/N ratios are much less sensitive to mixing processes, but the figure suggests that the observational results agree with the trend of the 
thermohaline-induced extra-mixing model or the $1^{st}$ dredge-up model, which are very similar. 
The model in which the thermohaline- and rotation-induced extra-mixing act together lies lower than the observational values that were determined in our work for
Collinder\,261 and Melotte\,66, as well as in previous studies of open clusters. 
 
This study shows that mixing processes in evolved giants should be further investigated, both observationally and 
theoretically. 

\begin{acknowledgements}
This research has made use of the WEBDA database (operated at the Department of Theoretical Physics and Astrophysics of 
the Masaryk University, Brno),  of SIMBAD (operated at CDS, Strasbourg), of VALD (\citealt{Kupka2000}), and of NASA’s 
Astrophysics Data System. Bertrand Plez (University of Montpellier II) and Guillermo Gonzalez  
(Washington State University) were particularly generous in providing us with 
atomic data for CN and C$_2$ molecules, respectively. This work was partly supported (AD, GT, \v{S}M) by the grant from 
the Research Council of Lithuania (MIP-082/2015). Partial support was received (AB, SR) from the Italian 
PRIN MIUR 2010-2011, project "The Chemical and Dynamical Evolution of the Milky Way and Local Group Galaxies".

\end{acknowledgements}

\end{document}